\documentclass[preprint,proceedings]{rmaa}

\SetYear{2016}
\SetConfTitle{XV Latin American Regional IAU Meeting LARIM}

\title{A simulation chain for the LAGO Space Weather Program}

\author{H. Asorey\altaffilmark{1}, L. A. N\'u\~nez\altaffilmark{2,3} and M. S\'uarez-Dur\'an\altaffilmark{2} for the LAGO Collaboration\altaffilmark{4}.
}

  \altaffiltext{1}{Laboratorio Detecci\'on de Part\'iculas y Radiaci\'on, Centro At\'omico Bariloche \& Instituto Balseiro, Bariloche, Argentina}

  \altaffiltext{2}{Escuela de F\'isica, Universidad Industrial de Santander, Bucaramanga, Colombia.}

  \altaffiltext{3}{Departamento de F\'isica, Universidad de Los Andes, M\'erida, Venezuela}

  \altaffiltext{4}{The full list of members and institutions at http://lagoproject.org; e-mail: lago@lagoproject.org}

\suppressfulladdresses

\listofauthors{H. Asorey}
\listofauthors{L. A. N\'{u}\~{n}ez}
\listofauthors{M. S\'uarez-Dur\'an}
\indexauthor{Asorey, H.}
\indexauthor{N\'u\~nez, L.~A.}
\indexauthor{Su\'arez-Dur\'an, M.}
\addkeyword{Space Weather}
\addkeyword{Geomagnetic Storms}
\addkeyword{Astroparticle Physics}

\begin{document}
\maketitle 

\boldabstract{
We simulate the expected variations of background flux at two
particular sites of the Latin American Giant Observatory (LAGO) 
and found that these fluxes are sensible to the latitude and that
neutrons and muons components, of cosmic rays, are affected due
to the variation of the geomagnetic field (GF).
}
The low energy Galactic Cosmic Ray (LEGCR) flux is modulated by
physical mechanisms with very different time scales: long-term
are associated with the solar cycle while short-term ones are 
produced by transient perturbations in the calm solar wind, 
causing rapid decreases in the galactic cosmic ray intensity 
following by a slow exponential-like recovery, known Forbush 
Decreases.

LAGO (Latin American Giant Observatory) is an
extended Astroparticle Observatory oriented to basic research on:
the Extreme Universe, Space Weather, and Atmospheric Radiation at 
ground level, with single or small arrays of particle detectors at
ground level, spanning over different sites covering a huge range 
of geomagnetic rigidity cutoffs and atmospheric absorption/reaction
levels\,\citep{AsoreyEtal2016a}. LAGO Space Weather program 
simulates and measures the LEGCR at ground level by Solar
modulation, exploiting the global distribution of the LAGO
detection network.

\begin{figure}[h]
  \centering
  \includegraphics[width=\columnwidth]{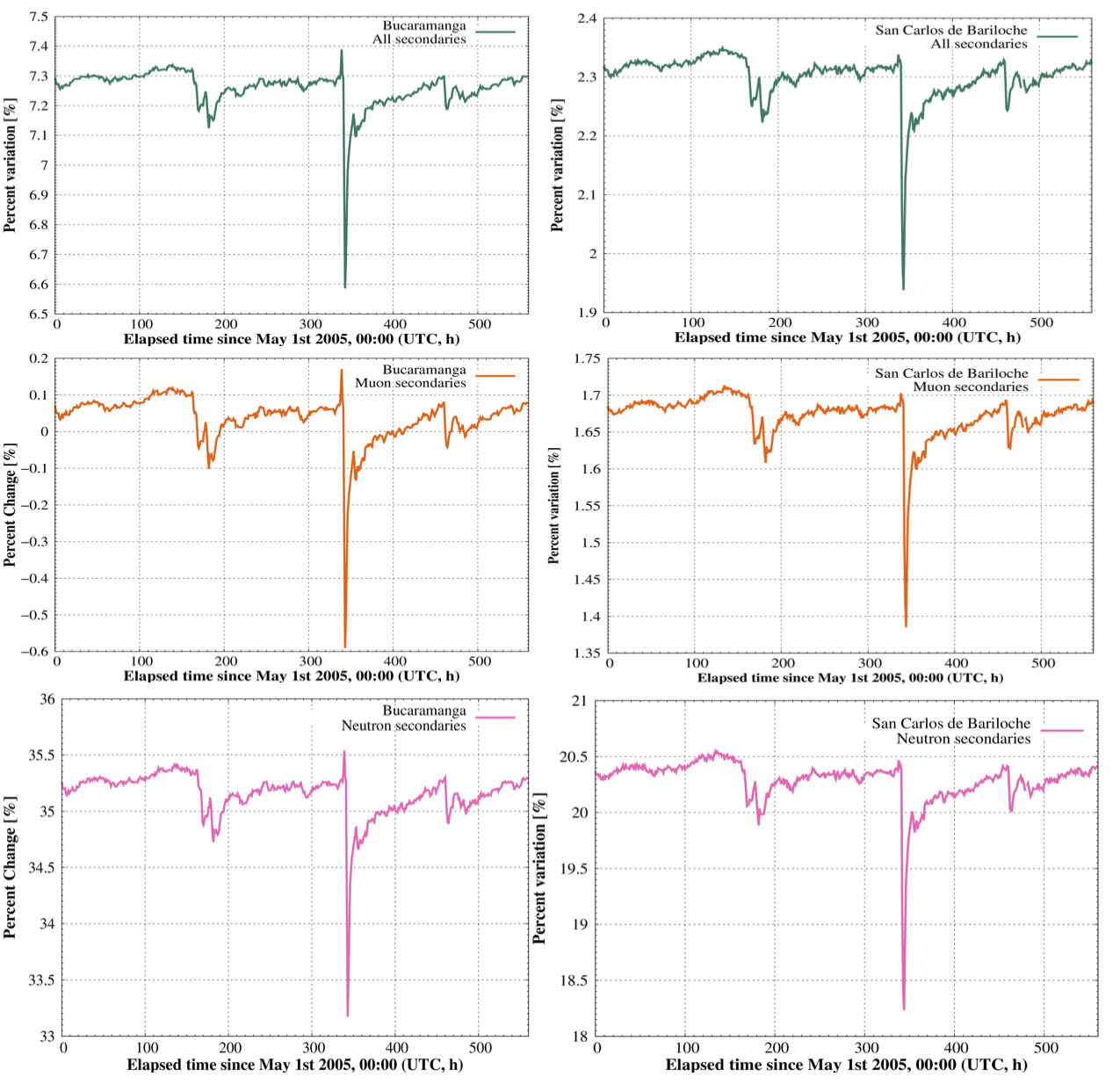}
	\caption{
		Simulations of the time evolution of the expected
		flux of secondaries during dynamic conditions of
		the GF, $\Delta \Xi_{1-2}$ for May of 2005 at
		Bucaramanga (left) and at San Carlos de Bariloche
		(right). The first row corresponds to the total
		flux of particles at ground level, while in the 
		second row we illustrates the evolution of the
		muon flux $\Delta\Xi_{1-2}^\mu$ and the third one
		displays the neutron flux $\Delta\Xi_{1-2}^n$. 
		There is a precise time coincidence of the
		simulated flux variation at both sites, which is 
		more significant in Bucaramanga than in
		Bariloche. It is also evident that the neutron
		flux at the ground level is the most affected 
		component by the GF activity.}
	\label{ForbushBMGSCB}
\end{figure}

The integrated particle flux and its modulation at the ground level 
are carefully simulated by considering local atmospheric profiles
and dynamic GF conditions \citep{AsoreyEtal2015B} at two particular 
LAGO sites: Bucaramanga, Colombia and San Carlos de Bariloche, 
Argentina. Our simulations show that the secondary flux is sensible 
to the latitude and that the secondary neutrons and muons at
ground level are affected flux components due to variations of GF 
during a space weather phenomenon (see Figure
\ref{ForbushBMGSCB}).\\
By combining the data measured at different locations of the LAGO
detection network, with those obtained from the detailed space
weather simulation chain, we are now capable to provide better 
understanding about the temporal evolution of the small and large
scales of solar and geomagnetic disturbances.

This project has been partially funded by VIE Universidad
Industrial de Santander.

\end{document}